\documentclass[twocolumn,runningheads]{svjour2}
\smartqed  
\usepackage{xspace}
\usepackage{graphicx}
\journalname{Astrophysics and Space Science}
\begin{document}

\title{An algorithm for solving the pulsar equation
}


\author{\L{}ukasz Bratek        \and
        Marcin Kolonko 
}


\institute{\L{}. Bratek \at
              H. Niewodnicza\'nski Institute for Nuclear Physics
              PAN\\
              ul. Radzikowskiego 152, 31-342 Krak\'ow, Poland\\
              Tel.: +48-12-6628284\\
              Fax: +48-12-6628458\\
              \email{bratek@ifj.edu.pl}           
           \and
            M. Kolonko \at
              H. Niewodnicza\'nski Institute for Nuclear Physics
              PAN \\
              ul. Radzikowskiego 152, 31-342 Krak\'ow, Poland\\
              Tel.: +48-12-6628064\\
              \email{kolonko@ifj.edu.pl}
}

\date{Received: date / Accepted: date}
\newcommand{\cckf}{Contopoulos, Kazanas \& Fendt\xspace}
\newcommand{\ckf}{\textsl{CKF}\xspace}
\newcommand{\ssw}{Scharlemann \& Wagoner\xspace}
\maketitle

\begin{abstract}
We present an algorithm of finding numerical solutions of pulsar
equation. The problem of finding the solutions was reduced to
finding expansion coefficients of the source term of the equation in
a base of orthogonal functions defined on the unit interval by
minimizing a multi-variable mismatch function defined on the light
cylinder. We applied  the algorithm to \ssw boundary conditions by
which a smooth solution is reconstructed that by construction passes
successfully the Gruzinov's test of the source function exponent.
\keywords{pulsars: general \and stars: neutron \and stars: rotation}
\PACS{96.50.sb \and 97.10.Kc \and 97.60.Gb}
\end{abstract}

\section{Introduction}
\label{intro:0} Pulsar equation describes structure of
electromagnetic fields and currents in magnetosphere of aligned
rotator. The structure is uniquely determined by a scalar function
$\Psi$ which is a solution of the equation
\begin{equation}
(1-\rho^2)\nabla^2\Psi-\frac{2}{\rho}\partial_{\rho}\Psi+F(\Psi)=0.
\label{psreq}
\end{equation}
The $\Psi$ and the unknown source function $F(\Psi)$ define
structure of electromagnetic fields and currents in pulsar
neighborhood. We use cylindrical coordinates $\rho$, $z$ and $\phi$
with the axis of rotation of the pulsar as the symmetry axis. To
solve the equation we assume \ssw conditions which we specify later
on. This however means, that we do not consider the particle
production issue nor we make any tests for the shape of emitting
regions. For simplicity we do not take into account return currents,
however they can be incorporated by adding delta function in the
source term.
\section{Assumptions of \ssw model}
To describe structure of magnetosphere \ssw assumed in \cite{sw73}
the following conditions
\begin{enumerate}
\item{{\bf Axial symmetry and stationarity}} \label{axl:2} The assumptions
follow naturally from the simplifying requirement that the axis of
rotation of a neutron star overlaps with its dipole momentum
(aligned rotator). The axis is therefore the axis of cylindrical
symmetry of the whole system and the system is independent of time.
The model is a particular case of the oblique rotator (\cite{k06}
and \cite{s05}). Consequently, the system does not generate the
pulsar effect because there is no modulation. However, we consider
the case of aligned rotator for simplicity. Such issue is
interesting, too, and it was discussed in many articles (like
\cite{ckf99},\cite{sw73}).

\item{{\bf Force-free approximation}} \label{ff:4} In this approximation one assumes
that inertial and gravitational forces are negligible, by which only
electric and magnetic forces are taken into account.  The assumption
agrees with considerations of Goldreich \& Julian \cite{gj69} who
showed that the electromagnetic interaction are stronger than
gravitational forces by a factor of $10^{8}$ for protons, and
$10^{11}$ for electrons. They also found that charge density just
above the neutron star surface cannot be zero. One can imagine
pulsar as extremely big atomic nucleus.

\item{{\bf Rigid co-rotation}} \label{rotation:5} Although we are aware of
the current works by Timokhin (\cite{t05} \cite{tim05}) who
considers models with critical point located inside the light
cylinder, we present our algorithm assuming that the point is
located just on the cylinder as in \ssw model. However, boundary
conditions can always be adapted to account for Timokhin
assumptions. Additionally, one may take into account return currents
but we do not consider them in our short paper aimed at presenting
our algorithm.

\item{{\bf Asymptotics}}
It is assumed that asymptotically $\Psi$ tends to the profile of the
split magnetic monopole of unit charge, while in the vicinity of the center a magnetic dipole should be located. The solution should
be smooth everywhere apart from the equatorial plane outside the
light cylinder.

\end{enumerate}
\section{Numerics}
\label{numr:6}
To find $\Psi$ with \ssw conditions in the whole physical
space we adapted pulsar equation as follows
\begin{enumerate}
\item{{\bf Compactification}}
is required to perform calculations on a finite size lattice.
For example, one can choose the mapping
\begin{equation}
x=\frac{2 \rho}{\sqrt{3+\rho^2}},\qquad y=\frac{2 z}{\sqrt{3+z^2}}.
\label{compact:18}
\end{equation}
that transforms the original infinite physical domain
$(\rho,z)\in[0,+\infty]\times[-\infty,+\infty]$ onto the square
$(x,y)\in[0,2]\times[-2,2]$. By symmetry it suffices to consider
only the region $y\ge0$. As an example in figure \ref{fig:1} there
are shown profiles of the monopole and of the dipole in the compactified domain.
\begin{figure*}
\centering
  \includegraphics[width=0.72\textwidth]{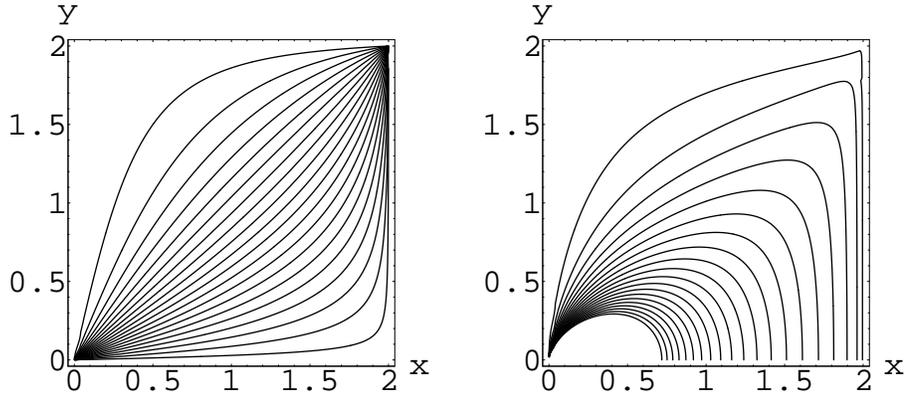}
\caption{The monopole and the dipole in the compactified domain for $y>0$.}\label{fig:1}       
\end{figure*}
\item{{\bf Boundary conditions}}
\label{bound:12} \ssw boundary conditions \cite{sw73} in the compactified domain
are given in table below \vskip 4mm
\begin{tabular}{ccc}
\hline\hline
$x$ & $y$ & $\Psi(x,y)=\Psi(x,-y)$ \\
\hline
$(1,2)$ & $0$ & $1$\\
$2$ & $(0,2)$ & $1$\\
$(0,2)$ & $2$ & $0$\\
$0$ & $(0,2)$ & $0$ \\
\hline
\hline
$x$ & $y$ & $\partial_y \Psi(x,y)$\\
\hline
$(0,1)$ & $0$ & 0\\
\hline \hline \label{tbl:17}
\end{tabular}
\vskip 2mm
In fact, we could use {\it any} boundary
conditions with our algorithm, in particular, the $Y$ point could be moved to the interior of the light cylinder.
\item{{\bf Discretization}}
\label{discrete:8}  Once we discretize our domain we may integrate
the pulsar equation. For a given $n$ the compactified lattice is
defined by nodal points of the grid $(j,k)$ such that $x=2j/(2n+1)$,
$y=k/n$, $j=0,1,\dots,2n+1$, $k=0,1,\dots 2n$ and
$U_{j,k}=\Psi(x,y)$, thus the singular line $x=1$ is not used during
integration. We thereby avoided the cumbersome problem of matching
solutions along the light cylinder and carried on calculations on a
single grid. The integration grid together with boundary conditions
is shown in figure \ref{fig:2}.
\item{{\bf The source function}}
The source function $F(\Psi)$ is nonzero in the domain
$\Psi=(0,1)$ (such normalization is possible), then in regions where $\Psi>1$ corotation takes place. Morover, from the analysis of pulsar equations with \ssw conditions it follows that $F'(0)=4$. Additionally, we take into account the result of paper by Gruzinov \cite{g05} that $F(\Psi)\sim(1-\Psi)^{\alpha}$ with $\alpha\approx7/12$ as $\Psi\to1^{-}$. Therefore, one may expand $F(\Psi)$ in the basis of the Jacobi polynomials $f_{i}(\Psi)=P^{7/6,4}_{i}(\Psi)$, $i=0,1,2,\dots$, where we used the convention that $F(\Psi)$ is the same as for the unit charge monopole at $c_i=0$, $i=0,1,2,\dots$.
\begin{equation}
F(\Psi)=2\Psi(1-\Psi)(2-\Psi)+\Sigma_i c_i\Psi^2(1-\Psi)^{\alpha}f_i(\Psi)
\label{srcfn:19}
\end{equation}
The key idea is to find the expansion coefficients such that $\Psi$
be smooth on the light cylinder. This can be done by minimizing an
error function $\mathcal{E}$ (defined below) measuring departure
from smoothness. By taking only a few initial terms of the expansion
the problem of finding a solution is reduced to finding the minimum
of $\mathcal{E}$ which is standard in numerical analysis. As an
aside, we remark that by adding a continuous representation of a
delta function $\delta_n$ to the above definition of $F(\Psi)$, that
is, by replacing $F(\Psi)$ with $F(\psi)+\beta\delta_{n}(1-\Psi)$,
on can similarly find a solution with return currents along
separatrix, where $\beta$ is the additional parameter to be
determined by the minimization.
\item{{\bf The error function}}
\label{error:9} To compute the error function we used the formula
\begin{equation}
\mathcal{E}(c_{-1},c_0,c_1,...,c_m)= \label{error:20}
\end{equation}
\begin{equation}
 =\sqrt{\sum_{k=0}^{2n-1}w_k \cdot \left[
\frac{U_{n+1,k}-U_{n,k}}{h_x}-\frac{2}{3}F \left(
\frac{U_{n+1,k}+U_{n,k}}{2} \right) \right]^2} \label{errorfn:16}
\end{equation}
which utilizes the original smoothness condition
$2\partial_{\rho}\Psi=F(\Psi)$ at $\rho=1$, and $w_k$ is arbitrary
discrete weight function. The $c_{-1}$ is proportional to the dipole
momentum of $\Psi$ and $c_0,c_1,\dots$ are the expansion
coefficients. The key point of our algorithm is, for a given $m$, to
find such a point $\{c_{-1},c_0,c_1,...,c_m\}$ at which
$\mathcal{E}(c_{-1},c_0,c_1,...,c_m)$ has the minimum. Finding
minima of multi-variable functions is a standard problem in
numerical analysis. It should be clear that the algorithm of finding
solutions differs qualitatively from the method presented by CKF in
\cite{ckf99}.
\end{enumerate}

\section{The results}
\begin{figure*}
\centering
  \includegraphics[width=0.65\textwidth]{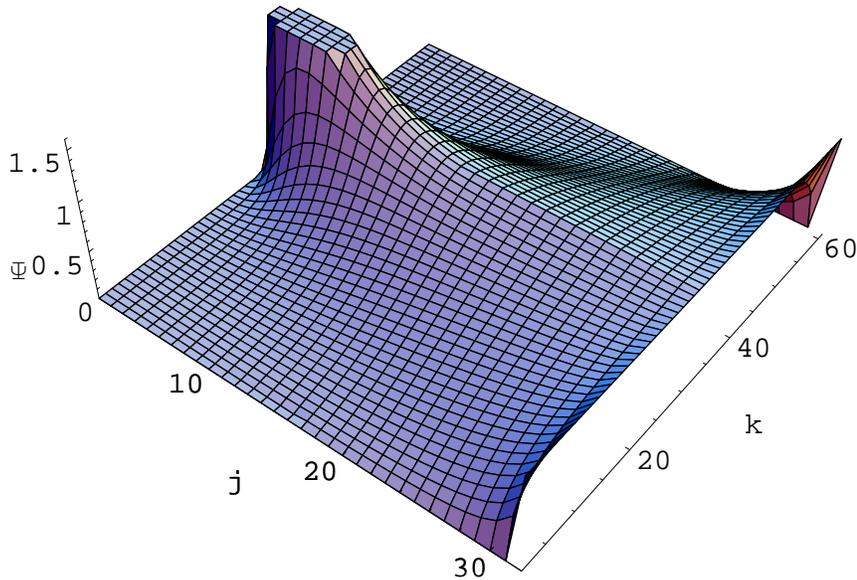}
\caption{The solution of pulsar equation with \ssw conditions shown in the compactified domain.}
\label{fig:2}       
\end{figure*}

In picture \ref{fig:2} and \ref{tab:2} are shown the whole profile
of $\Psi$ in the compactified domain and the source function,
respectively, which were found with the help of our algorithm for
$n=15$ (we used small $n$, where $n$ is the grid size, to show that
quite good results can be obtained with the help of our algorithm
even on very small size grids). In the table below \ref{tab:2} there
are shown also the corresponding parameters $c_0$ and
$c_1,\dots,c_6$ obtained for $n=15$. We remind we neglected in the
presentation the singular return currents to speed-up finding
solutions, but one can easily modify the expansion of $F(\Psi)$ to
account for it as discussed earlier.
\begin{table}[h]\label{tab:2}
\begin{tabular}{|c|r|r|}
\hline $i$& $c_i$& $\sigma_i$\\
\hline $0$& $10.498$ & $0.004$\\
$1$& $-0.1687$ & $0.0002$\\
$2$& $0.0740$ & $0.0005$\\
$3$& $-0.0266$ & $0.0005$\\
$4$& $0.0137$ & $0.0006$\\
$5$& $-0.0079$ & $0.0005$\\
$6$& $0.0072$ & $0.0002$\\
\hline
\end{tabular}
\caption{Location of the minimum of error function \ref{error:20} we
found for $n=15$ and $m=6$; $c_i$ are coordinates of the minimum and
$\sigma_i$ are uncertainties of the coordinates.}
\end{table}
\section{Discussion and summary}
\label{discuss:15} The new method of calculating source function
seems to work well. For bigger grids we don't presume they would
improve the results qualitatively. We can repeat our calculations
with Timokhin (\cite{tim05}) boundary conditions.

In our solution the surface of $\Psi$ is smooth apart from the equatorial plane outside the light cylinder.\\
The introduction to the problem can be found in
\cite{bkk06}.

%
\begin{figure*}
\centering
  \includegraphics[width=0.82\textwidth]{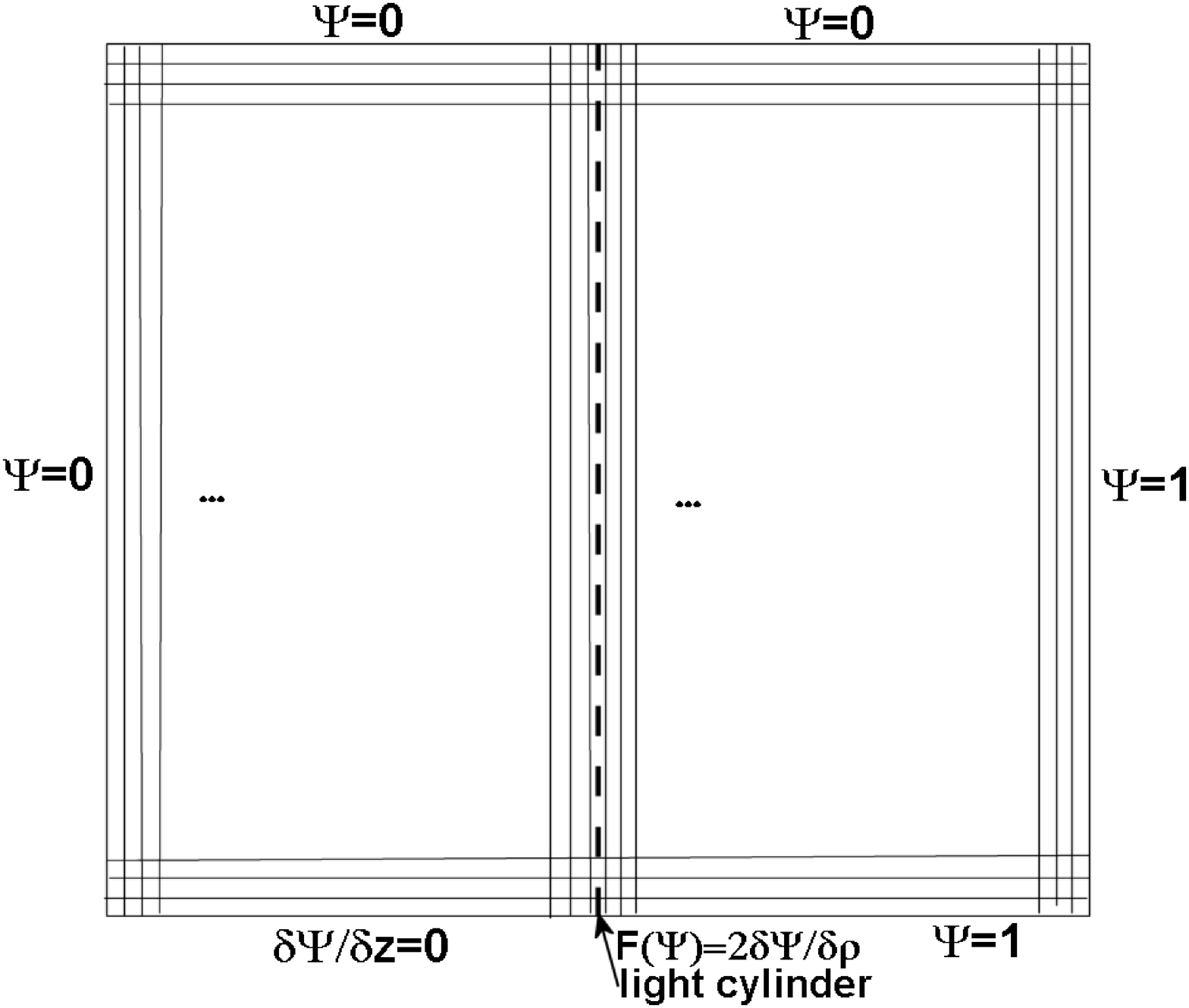}
\caption{Compactified integration grid which covers whole physical space. The light cylinder
which is singular surface of the pulsar equation is located between adjacent grid columns.}
\label{fig:3}       
\end{figure*}
\begin{figure}
\includegraphics[width=0.47\textwidth]{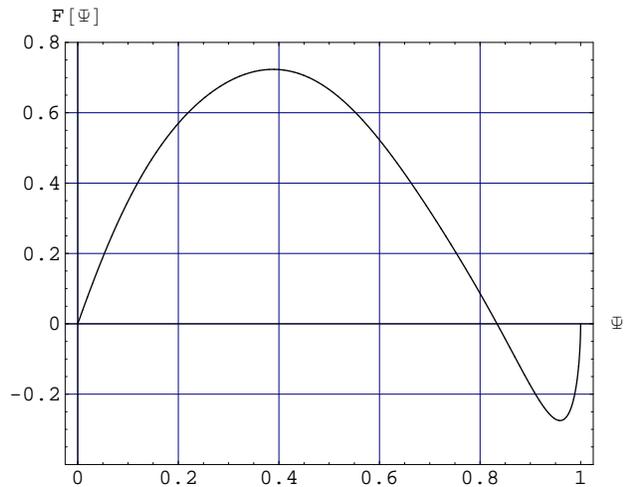}
\caption{Source function $F(\Psi)$ found for $n=15$ and
$m=6$.}\label{fig:msh15}
\end{figure}
%




\end{document}